\documentclass[12pt,a4paper]{article}
\pdfoutput=1
\usepackage{epsfig}
\usepackage{comment}
\usepackage{latexsym}
\usepackage{color}
\usepackage{amsmath}
\usepackage[english]{babel}

\newcommand{\mysquare}[0]{\raise-.2ex\hbox{{\Large$\Box$}}}
\def\lsim{\mathrel{\rlap {\raise.5ex\hbox{$ < $}}
{\lower.5ex\hbox{$\sim$}}}}
\def\gsim{\mathrel{\rlap {\raise.5ex\hbox{$ > $}}
{\lower.5ex\hbox{$\sim$}}}} \topmargin -1.5cm \textheight=22.5cm \textwidth=16.5cm
\catcode`\@=11
\newcount\hour
\newcount\minute
\newtoks\amorpm
\hour=\time\divide\hour by60 \minute=\time{\multiply\hour by60 \global\advance\minute by-\hour}
\edef\standardtime{{\ifnum\hour<12 \global\amorpm={am}%
        \else\global\amorpm={pm}\advance\hour by-12 \fi
        \ifnum\hour=0 \hour=12 \fi
        \number\hour:\ifnum\minute<10 0\fi\number\minute\the\amorpm}}
\edef\militarytime{\number\hour:\ifnum\minute<10 0\fi\number\minute}
\def\draftlabel#1{{\@bsphack\if@filesw {\let\thepage\relax
   \xdef\@gtempa{\write\@auxout{\string
      \newlabel{#1}{{\@currentlabel}{\thepage}}}}}\@gtempa
   \if@nobreak \ifvmode\nobreak\fi\fi\fi\@esphack}
        \gdef\@eqnlabel{#1}}
\def\@eqnlabel{}
\def\@vacuum{}
\def\draftmarginnote#1{\marginpar{\raggedright\scriptsize\tt#1}}
\def\draft{\oddsidemargin -.2truein
        \def\@oddfoot{\sl preliminary draft \hfil
        \rm\thepage\hfil\sl\today\quad\militarytime}
        \let\@evenfoot\@oddfoot \overfullrule 3pt
        \let\label=\draftlabel
        \let\marginnote=\draftmarginnote
   \def\@eqnnum{(\theequation)\rlap{\k

 ern\marginparsep\tt\@eqnlabel}%
\global\let\@eqnlabel\@vacuum}  }

\newcommand{\be}[0]{\begin{equation}}
\newcommand{\ee}[0]{\end{equation}}
\newcommand{\ba}[0]{\begin{eqnarray}}
\newcommand{\ea}[0]{\end{eqnarray}}
\newcommand{\di}{\displaystyle}
\def\bs{\begin{subequations}}
\def\es{\end{subequations}}

\def\thebibliography#1{%
\vskip 0.5cm \centerline{\bf \Large References}
\list{%
[\arabic{enumi}]}{\settowidth\labelwidth{[#1]} \leftmargin\labelwidth \advance\leftmargin\labelsep
\usecounter{enumi}}
\def\newblock{\hskip .11em plus .33em minus .07em}
\sloppy\clubpenalty4000\widowpenalty4000 \sfcode`\.=1000\relax}

\renewcommand{\theequation}{\arabic{section}.\arabic{equation}}

\renewcommand{\section}{\setcounter{equation}{0}\@startsection
{section}{1}{0mm}{-\baselineskip}{0.5\baselineskip} {\normalfont\Large\bfseries}}

\renewcommand{\subsection}{\@startsection
{subsection}{2}{0mm}{-\baselineskip}{0.5\baselineskip} {\normalfont\large\bfseries}}

\renewcommand{\subsubsection}{\@startsection
{subsubsection}{3}{0mm}{-\baselineskip}{0.5\baselineskip} {\normalfont\normalsize\slshape}}

\usepackage{amssymb,amsfonts}
\usepackage{graphicx}
\usepackage{cite}

\newcommand{\bea}{\begin{eqnarray}}
\newcommand{\eea}{\end{eqnarray}}

\newcommand{\Z}{\mathbb{Z}}
\newcommand{\Ka}{K{\"a}hler }
\renewcommand{\O}{{\cal O}}

\newcommand{\abs}{|}

\newcommand{\when}{\mbox{when}}
\renewcommand{\and}{\mbox{and}}

\newcommand{\F}{{\cal F}}
\newcommand{\N}{{\cal N}}

\newcommand{\T}{{\cal T}}

\newcommand{\U}{{\cal U}}

\newcommand{\Y}{{\cal Y}}

\renewcommand{\b}{\bar}
\newcommand{\h}{\hat}

\renewcommand{\t}{\tilde}

\def\ca{{\cal A}}

\def\cf{{\cal F}}

\def\ck{{\cal K}}

\def\cm{{\cal M}}
\def\cn{{\cal N}}

\def\car{{\cal R}}

\def\ct{{\cal T}}
\def\cu{{\cal U}}

\def\cw{{\cal W}}

\def\nnR{\nonumber\\}

\newcommand{\demi}{\frac{1}{2}}

\def\tdt{\tilde{\ct}}

\DeclareFontFamily{U}{rsfs}{}         
\DeclareFontShape{U}{rsfs}{m}{n}{<5> rsfs5 <6><7> rsfs7          %
  <8><9><10><10.95><12><14.4><17.28><20.74><24.88> rsfs10}{}     %
\DeclareMathAlphabet{\mathfs}{U}{rsfs}{m}{n}                     %
\newcommand{\mfs}[1]{\mathfs {#1}}                               %

\begin{document}
\begin{titlepage}
\begin{flushright}
CPHT-RR098.1111 \\ November 2011
\end{flushright}

\vspace*{2cm}

\begin{centering}

{\bf \Large MODULI STABILIZATION IN\\
\vspace{3mm}
EARLY SUPERSTRING COSMOLOGY}

\vspace{8mm}
 {\Large Lihui Liu}

\vspace{4mm}

Centre de Physique Th\'eorique, Ecole Polytechnique$^\dag$
\\
F--91128 Palaiseau cedex, France\\
{\em lihui.liu@cpht.polytechnique.fr}\\
\vspace{8mm}

{\it Based on a talk given at the ``The Seventh International Conference Quantum Theory and Symmetries'', Prague, Czech Republic, on August 7-13, 2011}

\vspace{8mm}

{\bf\Large Abstract}

\end{centering}

\noindent We study moduli stabilization by thermal effects in the cosmological context. The implementation of finite temperature, which spontaneously breaks supersymmetry, induces an effective potential at one loop level. At the points where extra massless states appear in the string spectrum, the potential develops local minima whose depth depends on the temperature. Moduli attracted to these points acquire dynamical masses which decrease with cosmological evolution. This makes the coherent scalar oscillations dilute before nucleosynthesis, and the cosmological moduli problem is avoided. In particular, we study the effective potential induced by a maximally supersymmetric heterotic string gas for spacetime dimension $D\geq 4$, and a gas of type II strings compactified on Calabi-Yau three-folds ($D=4$). In the former case, the local minima of the potential arise at enhanced gauge symmetry points, which can stabilize all moduli but the dilaton. In the latter case, the local minima are reached at the loci where 2-cycles or 3-cycles in the Calabi-Yau space shrink to zero size, accompanied with either conifold transitions or non Abelian gauge symmetries. This stabilizes the type II moduli which characterize the deformation of these shrinking cycles. Moduli stabilization in the dual string models is also investigated by heterotic/type I S-dualities and type II/heterotic dualities.

\vfill \hrule width 6.7cm \vskip.1mm{\small \small \small
\noindent
$^\dag$\ Unit{\'e} mixte du CNRS et de l'Ecole Polytechnique,
UMR 7644.}



\end{titlepage}

\section{Introduction and outline}

Moduli stabilization is a long standing issue in the superstring phenomenology. In fact, the presence of moduli fields in supersymmetric compactifications leads to difficulties: not only the massless scalars are in contradiction with observations of the gravitational force, but also, being continuous parameters in the couplings and mass spectrum, they imply a loss of predictability of the theory. Much attention is thus drawn to the search for mechanisms which attract the moduli fields to certain preferred values, where scalar masses are generated.

On the other hand, the scalar masses are subjected to constraints from cosmology. Basically, when the scalar fields oscillate coherently in the potential well, the energy of oscillation dominates the total energy of the universe \cite{Preskill:1982cy}, until the scalar particles decay. The productions of the decay can alter the primordial abundances of the light nuclei produced by nucleosynthesis. Also, the huge amount of entropy production during the decay can wash out the baryon number asymmetry. This is termed as the cosmological moduli problem, which was initially identified in the framework of supersymmetric standard models \cite{CosmModProb,Ellis:1986zt,Moroi:1994rs}. One applaudable solution to these is to require the scalar masses be of $O(10)$TeV order. It is pointed out in \cite{Ellis:1986zt} that once this is satisfied, the decay of these scalar particles reheats the universe to a temperature of order $1$MeV, high enough to restart the nucleosynthesis. Then it is found in \cite{Moroi:1994rs} that the baryon number asymmetry can also be saved by the $O(10)$TeV order scalar mass if the baryogenesis is due to the Affleck-Dine mechanism \cite{Affleck:1984fy}.

Here I present our recent work \cite{Estes:2011iw,Liu:2011nw} where the above problems were addressed by investigating thermal string effects. It was shown in \cite{Patil:2004zp} that a gas of string modes, carrying both winding and momenta, can generate a free energy that enables stabilization of radii moduli. A quantum version of this effect has been presented in \cite{cosmoD, Angelantonj:2006ut,Estes:2011iw,Liu:2011nw}, with the thermal gas and free energy replaced by virtual strings which induce an effective potential. To avoid generating a large cosmological constant, the cosmology is addressed in the context of no-scale models \cite{Noscale}. The latter are defined at classical level by backgrounds associated to vanishing minima of the scalar potential, with flat directions parameterized by the spontaneous supersymmetry breaking scale.

For simplicity, we consider here only temperature breaking of supersymmetry. At the level of conformal field theory on the worldsheet, the implementation of finite temperature amounts to a Scherk-Schwarz reduction on the Euclidean time circle of radius $R_0$, with boundary conditions associated to the spacetime fermion number \cite{Kounnas:1989dk}. The string frame temperature is $\h T=\h \beta^{-1}=1\big/2\pi R_0$ and the Einstein frame temperature is $T=e^{{2\over D-2}\phi^{(D)}}\, \h T$, where $\phi^{(D)}$ is the $D$-dimensional dilaton. The supersymmetry is thus broken spontaneously at the scale $T$. We restrict our attention to the intermediate era between the Hagedorn phase transition \cite{Florakis:2010is} and the electroweak phase transition, so that  $M_{\rm string}\gg T\gg \Lambda_{\rm EW}$.

To build phenomenologically viable models however, it is necessary to also include zero temperature spontaneous supersymmetry breaking. Otherwise as the temperature drops during the cosmological evolution, the supersymmetry broken by temperature will be restored. The case with another Scherk-Schwarz reduction performed in one of the internal dimensions is intensively studied in Refs.\cite{cosmoD}, where it is shown that the supersymmetry breaking scale $M_{\rm SUSY}$ induced in this internal dimension evolves proportionally with $T$. It is expected that by the end of the intermediate era, when $T$ approaches $\Lambda_{\rm EW}$, the radiative corrections induced by infrared effects start to destabilize the Higgs potential, freezing $M_{\rm SUSY}$ at about $O(1)$TeV order. This gives an account of the hierarchy $M_{\rm SUSY}\ll M_{\rm Plank}$ .

The breaking of supersymmetry generates a nontrivial vacuum-vacuum amplitude, which we compute at one-loop level, supposing that the string theory is at weak coupling. This amplitude is just the thermal partition function of the string gas computed up to one-loop level, which we denote by $Z$. It gives rise to the free energy density by $\cf=-{Z\over \beta V}$, ($V$ the space volume in Einstein frame). The back-reaction of $\cf$ on the spacetime background is dictated by the one-loop effective action
\begin{align}
    S=\int d^Dx\sqrt{-g}\Big[ {R\over 2} -\demi F_{MN}\partial \Phi^M \partial \Phi^N - \cf\big(T, \vec \Phi\big)\Big], \label{fed1}
\end{align}
where $\Phi^M$ are the moduli, and the metric components $F_{MN}$ are functions of these moduli. Since $\cf$ appears in the action as the effective potential, moduli attractors should be its local minima.

It can be shown that when we only have temperature breaking of supersymmetry, the free energy density takes the form \cite{Estes:2011iw}:
\begin{align}
{\cal F}(T,&\vec\Phi) = -\int_0^\infty \frac{d\ell}{2\ell} \frac{1}{(2\pi \ell)^{\frac{D}{2}}} \sum_s e^{- {1\over2} M_{\!s} (\vec\Phi)^2 \ell} \sum_{k_0 \in \mathbb Z} e^{-\frac{ (2k_0+1)^2}{2 T^2 \ell}} = -T^D\sum_s G \big(M_{\!s}(\vec\Phi)/T \big), \label{fed2}
\end{align}
where $M_s$ is the tree-level mass of the $s$-th string state, which can depend on the moduli. The function $G(x)$ is defined in terms of the modified Bessel function of the second kind (see \cite{Estes:2011iw} for more details). It peaks at $x=0$ and is exponentially suppressed at large $x$. Therefore only light states give significant contribution to $\cf$. By consequence the local minima of $\cf$ appear at the vacuum expectation values (VEV's) of $\vec \Phi$ where some massive states in the string spectrum become massless. These states can originate either from the perturbative spectrum or from non perturbative objects such as D-branes. String-string dualities can help figure out the non perturbative contribution.

The local minima of $\cf$ induce time-dependent scalar masses, instead of constant ones. This ensures that the universe is radiation dominated at the exit of the intermediate era, which is crucial to the resolution of the cosmological moduli problem. In order to show this, we take the flat Robinson-Walker metric $ds^2=-dt^2+a(t)^2 d\vec x^2$ (in the Einstein frame). Solving the equations of motion about a local minimum of $\F$, say $\vec \Phi_0$, we obtain the following time evolution of the scale factor, the temperature, and the total energy density:
\begin{align}
\label{fed8}
    a(t)\propto 1/T(t) \propto t^{2/D},\ \ \rho_{\rm tot}\propto H^2 \propto a^{-D}.
\end{align}
The coherent moduli field oscillations obey the equation
\begin{align}\label{fed9}
    \ddot\epsilon^M+(D-1)H\dot\epsilon^M+\Lambda^M_{\ N}\epsilon^N=0,
\end{align}
where we let $\vec \Phi=\vec \Phi_0+ \vec \epsilon$, and we have the squared-mass matrix $\Lambda^M_{\ N}=\big(F^{MP}\F_{PN}\big)_{\vec\Phi_0}$, with $\F_{PN}:=\partial^2 \F\big/ \partial \Phi^P\Phi^N$, and $F^{MN}$ the inverse of $F_{MN}$. Using Eq.(\ref{fed2}) one can show that $\Lambda^M_{\ N}\propto T^{D-2}$. Thus with Eq.(\ref{fed9}) we have the usual scalar dynamics, but with scalar masses depending on the temperature, hence on time. This results in the late-time scalar oscillation behavior $\epsilon\sim t^{-1/2}\sin\big(\lambda\, t^{2/D}+{\rm phase}\big)$, instead of $t^{1/D-1}\sin\big(\lambda\, t+{\rm phase}\big)$ for constant mass, where $\lambda^2$ is some eigenvalue of the squared-mass matrix. Therefore the energy density stored in the scalar oscillations behaves asymptotically as
\begin{align}
    \rho_{\Phi}=\demi F_{MN}\big|_{\vec\Phi_0}\,\, \dot \epsilon^M\dot \epsilon^N\sim t^{{4\over D}-3}\propto a^{2-3D/2}, \label{fed10}
\end{align}
while the result for the case of constant scalar mass is $\rho_{\Phi}\sim a^{1-D}$ which obviously dominates over the radiation energy $\rho_{\rm rad}\sim a^{-D}$ for any spacetime dimension. Back to the case of dynamical mass, where Eq.(\ref{fed10}) holds, when $D\geq 5$, the universe is radiation dominated, since compared to Eq.(\ref{fed8}), we have $\rho_{\Phi}\ll \rho_{\rm tot}$. However $D=4$ is a marginal case where even though the metric evolution appears as that of a radiation dominated universe ($\rho_{\rm tot}\propto H^2 \propto a^{-4}$), the energy of coherent scalar oscillations is not dominated. Instead, it takes up a constant portion in the total energy ($\rho_{\rm tot}\propto \rho_{\Phi} \propto a^{-4}$). This is due to the over-simplified supersymmetry breaking mechanism that we adopt here. It is shown in \cite{cosmoD} that the $D=4$ case is also radiation dominated once an additional source of spontaneous supersymmetry breaking is introduced in the internal space.

In the following, we investigate the cosmology induced by two specific string models: the maximally supersymmetric heterotic strings and the Calabi-Yau (CY) compactification of type II strings.

\section{Heterotic cosmology and type I dual}\label{hetTI}

We start with the cosmology induced by weakly coupled $SO(32)$ heterotic strings compactified on a factorized torus $\prod_{i=D}^9 S^1(R_{{\rm h}i})$, where the subscript h indicates heterotic quantities. The model have maximal number of supersymmetry, so that the metric $(F_{MN})$ in Eq.(\ref{fed1}) is exact at tree level. Let the moduli space be coordinatized by the $D$-dimensional dilaton $\phi_{\rm h}^{(D)}:=\phi_{\rm h}^{(10)}-{1\over 2}\sum_{i=D}^9 \ln(2\pi R_{{\rm h}i})$ and all the internal radii $R_{{\rm h}i}$ with $i=D,\dots,9$. Computing the thermal one-loop amplitude, we find that when all radii satisfy $\abs R_{{\rm h}i}-1/R_{{\rm h}i} \abs < 1/(2\pi R_{{\rm h}0})$, $i=D,...,9$, the corresponding free energy density takes the form  \cite{Estes:2011iw}:
\begin{align}
\label{HTI2}
\F_{\rm h}= -T^D \left\{ n_0\, c_D+ \sum_{i=D}^9 n_1\, G\bigg(2\pi R_{{\rm h}0}\Big\abs {1\over R_{{\rm h}i}}-R_{{\rm h}i}\Big\abs\bigg)+\O(e^{-2\pi R_{{\rm h}0}})\right\},
\end{align}
where the coefficients $n_0$ and $n_1$ are positive, associated to the counting of states. The first term in the above expression is from massless states. The second term involving the $G$-function shows that $\cf_{\rm h}$ reaches a local minimum at the self T-dual point $R_{{\rm h}i}=1$ ($i=D,\dots,9$), due to the states of masses $\big|{1\over R_{{\rm h}i}}-R_{{\rm h}i}\big|$. These are just the non Cartan components responsible for the gauge symmetry enhancement $U(1)\rightarrow SU(2)$ in each internal circle. In fact in heterotic strings, the correspondence between the enhancement of gauge symmetry and the local extrema of the free energy is true to all loop levels \cite{Ginsparg:1986wr}. Therefore the internal radii can all be stabilized at the value $1$ where we have $SU(2)^{10-D}$ enhanced symmetry. Moreover for $D\ge 5$, the string coupling $\lambda_{\rm h}^{(D)}=e^{\phi_{\rm h}^{(D)}}$ freezes on the flat direction to some constant value determined by the initial conditions. For $D=4$, the dilaton $\phi_{\rm h}^{(4)}$ does not converge to a constant but instead decreases logarithmically with the cosmological time.

We switch to the dual type I picture. If we perform naive perturbative computation $Z_{\rm I}=\ct+\ck+\ca+\cm$ to obtain the free energy density, we will find no local minimum of $\cf_{\rm I}$, since no perturbative effect can lead to gauge symmetry enhancements in maximally supersymmetric type I strings. We thus seek to include non perturbative effects which can be inferred from heterotic strings through string-string S-dualities. In dimension $D$, the duality dictionary for Einstein frame quantities is \cite{allD}
\begin{align}
\label{HTI3}
\begin{array}{l}
\di R_{{\rm h}i}={R_{{\rm I}i}\over \sqrt{\lambda_{\rm I}}}\equiv R_{{\rm I}i}\, {e^{-{1\over 2}\phi_{\rm I}^{(D)}}\over \left(\prod_{j=D}^92\pi R_{{\rm I}j}\right)^{1/4}}\; ,\quad i=0 \mbox{ or } D,...,9,\\ \di
\phi_{\rm h}^{(D)}=-{D-6\over 4}\, \phi_{\rm I}^{(D)}-{D-2\over 8}\sum_{i=D}^9\ln\left(2\pi R_{{\rm I}i}\right),
\end{array}
\end{align}
where $\lambda_{\rm I}$ is the type I string coupling in ten dimensions. When applying this duality map, the heterotic states that induce the local minimum in Eq.(\ref{HTI2}) are sent to non perturbative states of masses $\left| {1\over R_{{\rm h}i}}-{R_{{\rm h}i}\over \lambda_{\rm I}}\right|$ on the type I side. From the type I point of view, they have the natural interpretation as D (or anti-D)-strings wrapped once along the circles $S^1(R_{{\rm I}i})$, with one unit of momentum. Therefore when all radii satisfy $\left| {1\over R_{{\rm I}i}}-{R_{{\rm I}i}\over \lambda_{\rm I}}\right| < {1\over 2\pi R_{{\rm I}0}}$, they are attracted to $R_{{\rm I}i}=\sqrt{\lambda_{\rm I}}$, where we have the enhanced gauge symmetry $SU(2)^{10-D}$ due to D-string states. The type I dilaton freezes somewhere along its flat direction just as its heterotic dual except for $D=6$ where it is stabilized while the internal space volume $\prod_{i=D}^9(2\pi R_{{\rm I}i})$ freezes along a flat direction. This is because in $D=6$ the duality map Eq.(\ref{HTI3}) exchanges internal volumes and string couplings. Another subtlety arising from Eq.(\ref{HTI3}) is that, since the heterotic theory is always in the weak coupling regime, the type I dual is strongly coupled for $D>6$ and weakly coupled for $D<6$. However our result is still valid at small coupling for $D>6$ since the D-string states, responsible for the stabilization of $R_{{\rm I}i}$, are BPS states whose masses are protected by supersymmetry.

The D-string state contribution can also have an E1-instanton interpretation, following the lines of Refs.\cite{Bachas:1997mc}. For simplicity, we consider the compactification on $S^1(R_{{\rm I}9})$. This contrasts the zero temperature case where E1-instantons arise for $D\leq 8$. Starting from the heterotic side, we can easily express the thermal partition function as a sum over worldsheet instantons. When sending this heterotic result to the type I side using the dictionary (\ref{HTI3}), the corresponding type I partition function contains a sum of E1-instantons, which is explicitly \cite{Estes:2011iw}
\begin{align}
\label{ZndinsI}
Z_{\rm I}^{E1}\!=\!{\h V_{\rm I}^{(10)}\over  (2\pi)^{10}} \; 2\!\!\!\!\!\!\!\!\sum_{\rm \scriptsize E1 \, instantons}\!\!\!\!\!\!\!\!s_0\, {e^{\frac{2i\pi}{\lambda_{\rm I}}\Upsilon_{\rm I}}\over \Upsilon_{{\rm I}2}\, \Y_{{\rm I}2}^4} \sum_{n=0}^4\!\!\left[\!{\alpha_n\over (2\pi \Upsilon_{{\rm I}2})^n} \!\! \sum_{\b A\ge -1}\!\!b_{\b A}\!\left(\!\frac{1}{\lambda_{\rm I}}+\b A\, {\Y_{{\rm I}2}\over \Upsilon_{{\rm I}2}}\!\right)^{4-n}\!\!\!\!e^{2i\pi \Y_{\rm I}\b A}\!\right]\!+c.c. +\O(e^{-4\pi {R_{{\rm I}0}\over \sqrt{\lambda_{\rm I}}}}),
\end{align}
with the \Ka and complex structure moduli $\Upsilon_{\rm I}$ and $\Y_{\rm I}$ of the torus $S^1(R_{{\rm I}0})\times S^1(R_{{\rm I}9})$
\begin{align}
    \left\{\begin{array}{l}\Upsilon_{\rm I}=i\Upsilon_{{\rm I}2}=i(2\t k_0+1)R_{{\rm I}0}\cdot n^9R_{{\rm I}9}\\\di \Y_{\rm I}=\Y_{{\rm I}1}+i\Y_{{\rm I}2}={\t m_9\over n^9}+i\, {(2\t k_0+1) R_{{\rm I}0}\over n^9R_{{\rm I}9}}
\end{array}\right. ,\ \ \ \  n^9> \t m_9\ge 0\; , \; \,\t k_0\ge 0.
\end{align}
This result suggests it possible to derive from a pure type I point of view the free energy responsible for the stabilization of the internal moduli.

We can further consider generic toroidal compactifications, where all possible moduli are switched on. On the heterotic side, these moduli include the dilaton $\phi_{\rm h}^{(D)}$, the internal metric $g^{({\rm h})}_{ij}$, the internal antisymmetric tensor $B^{({\rm h})}_{ij}$, and the Wilson lines $Y^I_{({\rm h})i}$, where $i,j=D,\dots,9$ and $I=10,\dots,25$. Again, all moduli except the dilaton are attracted to the values associated to some enhanced gauge symmetry, where $\cf_{\rm h}$ is minimized locally. In the dual type I picture, moduli stabilization is inferred from the heterotic side through the dictionary
\begin{align}\label{HTI5}
\begin{array}{l}
    \di \phi_{\rm h}^{(D)}=-{D-6 \over4}\phi_{\rm I}^{(D)}-{D-2\over 8}\ln\sqrt{g^{(h)}}, \\ \di
    g^{({\rm h})}_{ij}={g^{({\rm I})}_{ij}\over \lambda_{\rm I}},\ \ \  B^{({\rm h})}_{ij}=C_{ij},\ \ \ Y^I_{({\rm h})i}=Y^I_{({\rm I})i}.
\end{array}
\end{align}
where $C_{ij}$ is the Ramond-Ramond 2-form. The subtlety is that now the dual type I moduli are stabilized by either non perturbative D-string states in the closed string sector or perturbative states in the open string sector. For $D\neq6$, all type I moduli are stabilized except the dilaton, and for $D=6$ however, the dilaton is stabilized while the internal volume freezes on a flat direction.

As an explicit example, we examine the case of compactification on $T^2$, where we have on the heterotic side, the moduli $\T=B_{89}+i\sqrt{\h g_{88}\h g_{99}-\h g_{89}^2}$, $\U=\big(\h g_{89}+i\sqrt{\h g_{88}\h g_{99}-\h g_{89}^2}\big)/\h g_{88}$ and the Wilson lines $Y_i^I$ ($i,j=8,9$; $I=10,11,\dots,25$). The mass formula for perturbative F-string states is
\begin{align}
\label{HTI6}
    \!\! \hat{M}^2_{A,\vec{m},\vec{n},\vec{Q}}(\ct,\cu,Y)\! =\! \frac{1}{\ct_2\cu_2}\left|-m_8 \cu \!+m_9\! +\tdt n^8 \! +\Big(\tdt \cu-\demi \cw^I \cw^I\Big)n^9\! +\cw^IQ^I\right|^2 \!\! +\! 4A,
\end{align}
where $\cw^I:=\cu Y^I_8-Y^I_9$ and $\tdt:=\ct+\demi Y^I_8\cw^I$, $\vec m$, $\vec n$ are the internal momenta and winding numbers, and $Q^I$ the root vector of the internal lattice $\Gamma_{O(32)/{\Z_2}}$. Using the mass formula we can figure out moduli attractors where there are states becoming massless. The enhanced gauge group can be determined from the Narrain lattice formed by the right-moving internal momenta of these states. For example we have the local attractor with $SU(3) \times SO(32)$ enhanced symmetry, where the moduli are stabilized at $Y^I_i=0$, $\ct=\cu=\demi+i\frac{\sqrt{3}}{2}$. Another less trivial example is the attractor with $SU(2) \times SO(34)$ enhanced symmetry, where the moduli are attracted to $\ct=\cu=i/\sqrt{2}$, $Y^{I\ge 10}_8=0$ and $Y^{10}_9=-Y^{11}_9 = -Y^{12}_9 = \dots =- Y^{25}_9 =-1/2$. In the dual type I picture the moduli stabilization follows from the dictionary (\ref{HTI5}).

\section{Type II cosmology and heterotic dual}\label{CY}

We turn to models with less supersymmetry. We consider cosmology in type II strings compactified on a Calabi-Yau (CY) three-fold $M$ of Hodge numbers $(h_{11}, h_{12})$. The moduli space is a Cartesian product $\cm_V\times \cm_H$. The vector multiplet moduli space $\cm_V$ of complex dimension $h_{11}$ is a special \Ka manifold, which is exact at tree level, because the dilaton lives in a hypermultiplet. The hypermultiplet moduli space $\cm_H$ of real dimension $4(h_{12}+1)$ is a quaternionic manifold, which contains the universal hypermultiplet accommodating the dilaton. Therefore $\cm_H$ is subjected to perturbative and non perturbative corrections. When $M$ is a $K3$ fibration, a dual heterotic string theory can exist, compactified on $K3\times T^2$. The string-string duality sends the type II vector multiplet moduli to the heterotic vector multiplet moduli and the same is true of the hypermultiplet moduli. Thus the stabilization of heterotic moduli can be inferred from the stabilization of the dual type II moduli.

On the type II side, the moduli space develops singular loci when the internal CY space undergoes extremal transitions. At these loci, some 2-cycles or 3-cycles in the CY space shrink to zero size, giving rise to a singular three-fold. This can lead to conifold transitions or non Abelian gauge symmetries, with extra massless states appearing in the low-energy spectrum. Nonsingular CY three-folds can be recovered by restoring the shrinking 2-cycles or 3-cycles to finite size. In the following analysis, we adopt the type IIA description, and suppose that the desingularization by restoring 2-cycles is always available. Indeed only in this case can we write down the effective gauge theory, following the analysis in \cite{Strominger:1995cz, Katz:1996ht}.

At the conifold locus, let $R$ 2-cycles in the CY space $M$, spanning an $S$-dimensional subspace of homology, shrink to separated nodes. Locally, $R$ monopole states become massless, described by $R$ hypermultiplets charged under $S$ $U(1)$-vector multiplets \cite{Strominger:1995cz}. When $R>S$, we can deform the shrinking 2-cycles into 3-cycles and obtain a topologically different CY space $M'$. The change in Hodge numbers is
\begin{align}
    h_{11}(M')= h_{11}(M)-S, \ \ h_{12}(M')= h_{12}(M)+R-S. \label{TII1}
\end{align}
Near the non Abelian locus, $N-1$ homologically independent 2-cycles, with the intersection matrix the Cartan matrix of $A_{N-1}$, shrink to zero size along a smooth curve $C$ of genus $g$. By the arguments in \cite{Katz:1996ht}, $N^2-N$ vector multiplets and $g(N^2-N)$ hypermultiplets become massless, giving rise to the gauge symmetry enhancement $U(1)^{N-1}\rightarrow SU(N)$ with $g$ hypermultiplets transforming in its adjoint representation. When $g>1$, we can construct a topologically different CY space $M''$ by deforming all shrinking 2-cycles into 3-cycles. The change in Hodge numbers is
\begin{align}
    h_{11}(M'')= h_{11}(M)-(N-1), \ \ h_{12}(M'')= h_{12}(M)+(g-1)(N^2-N)-(N-1). \label{TII2}
\end{align}
In both cases, the low energy effective theory about the singular loci containing all light fields is described by a gauged $\cn_4=2$ supergravity theory. Desingularizing the CY space by restoring the shrinking 2-cycles (3-cycles) corresponds to sitting in the Coulomb (Higgs) branch of the gauge theory. Therefore by our setup, the Coulomb branch must exist. The scalar fields in the light vector multiplets span a special \Ka manifold which contains $\cm_V$, and we denote its special coordinates by $\{X^I\}$, $I=1,\dots,n_V$. The scalar fields in the light hypermultiplets span a quaternionic manifold which contains $\cm_H$, and we let its real coordinates be $\{q^{\Xi}\}$, $\Xi=1,\dots,4n_H$. Here $n_V\geq h_{11}$ and $n_H\geq h_{12}+1$ are respectively the total number of light vector multiplets and light hypermultiplets. These scalar fields are divided into two groups: those participating in the extremal transition whose VEV's characterize the deformation of the vanishing cycles, and the rest which are spectators to the extremal transition. We then let $g_{I\!\b J}=g_{I\!\b J}(X^K)$ and $h_{\Lambda\Sigma} = h_{\Lambda \Sigma} (q^{\Xi})$ be the special \Ka metric and quaternionic metric. Due to the gauging, a scalar potential is generated. The supergravity action is now regular in the neighborhood of singular loci since the inclusion of all light states repairs the IR divergences.

\subsubsection*{Attraction to conifold transition loci}

Near the conifold locus, the scalar fields participating in the extremal transition are those in the vector multiplets of $U(1)^S$, $X^i$ ($i=1,\dots,S$), and those in the $R$ hypermultiplets charged under $U(1)^S$, $q^{\ca u}$ $(\ca=1,\dots,R;\ u=1,2,3,4)$. The conifold locus can be represented by $X^i=0=q^{\ca u}$ with suitable choice of parametrization. For simplicity we switch off the spectator scalar fields. In the neighborhood of the conifold locus, by performing power expansion in $X^i$ and $q^{\ca u}$ and imposing $U(1)^S$-isometry, we obtain the scalar part of the supergravity action to the lowest order \cite{Liu:2011nw}:
\begin{align}
    S=&{1\over \kappa^2_{(4)}}\int d^4x\,\sqrt{-g}\,\Big[\demi R- g_{i \b \jmath}\,\partial X^i\partial \b X^j-\nabla \mfs Q^{\ca\dagger}\nabla \mfs Q^{\ca} \nnR &- g_c^2\,e^{\ck_V}\, \sum_{\substack{i,j}}\Big(4\, \sum_{\ca}\, Q^{\ca}_i Q^{\ca}_j\, \b X^i X^j\, \mfs Q^{\ca\dagger}\mfs Q^{\ca}+ g^{i\b \jmath}\,\vec D_i \! \cdot\! \vec D_j \Big)\Big], \label{TII4}
\end{align}
where $Q^{\ca}_i$ is the charge of the $\ca$-th monopole under the $i$-th $U(1)$, $g_c$ the gauge coupling constant. The \Ka potential $\ck_V$, the special \Ka metric and the quaternionic metric in the above action are constant, taking their values at the conifold locus. Also we have defined the $SU(2)_{\car}$ doublet and the D-term:
\begin{align}
    \mfs Q^{\ca}=\left(\!\!\begin{array}{c} -q^{\ca 2}+i\,q^{\ca 1}\\ q^{\ca 3}+i\,q^{\ca 4}\end{array}\!\!\right),\ \ \ \vec D_i=\sum_{\ca}Q^{\ca}_i \mfs Q^{\ca \dagger} \vec \sigma \mfs Q^{\ca}.\label{TII5}
\end{align}
The action (\ref{TII4}) describes an $\cn_4=2$ supersymmetric Abelian gauge field theory formally coupled to gravity. We show that moduli are attracted to the conifold locus whether starting in the Coulomb branch or the Higgs branch.

\noindent $\bullet$ In the Coulomb branch, corresponding to the compactification on $M$, $X^i$ ($i=1,\dots,S$) obtain nonzero VEV's, while $q^{\ca u}$ ($\ca=1,\dots,R;\ u=1,2,3,4$) have zero VEV. Thus the VEV's of $X^i$ form $S$ of the $h_{11}(M)$ \Ka moduli, parameterizing the Coulomb branch vacua together with the moduli fields which are spectators to the conifold transition. The free energy density is \cite{Liu:2011nw}
\begin{align}
    \cf=-T^4\Big[n_0+\sum_s n_s G\Big({M_s\over T}\Big)\Big]+{\cal O}\big( e^{-{M_{\rm min}\over T}}\big), \label{TII501}
\end{align}
where $n_0$ and $n_s$ count respectively the massless states and the light monopole states. Also $M_{\rm min}$ is the minimum mass of the states which never become massless in the neighborhood of the conifold locus. We let the temperature be much lower than this mass, $T\ll M_{\rm min}$, so that the contribution from massive states is exponentially supressed. In the argument of the $G$-function, $M_s$ is the tree-level mass of the $s$-th light monopole state, which has the behavior $M_s\sim {\cal O}(X^i)$. Therefore at the conifold locus where $X^i=0$, the free energy density reaches its local minimum. Thus the $S$ \Ka moduli $X^i$ are attracted to the conifold locus.

\noindent $\bullet$ In the Higgs branch, corresponding to the compactification on $M'$, $q^{\ca u}$ ($\ca=1,\dots,R;\ u=1,2,3,4$) have nonzero VEV's subjected to the constraints $\vec D_i=0$ modulo gauge orbits, so that they parameterize $R-S$ of the $h_{12}(M')+1$ quaternionic directions in the complex structure moduli space. On the other hand, the VEV's of $X^i$ vanish, and the vector multiplets containing $X^i$ absorb $S$ hypermultiplets to form $S$ long massive vector multiplets. The free energy density takes the same form of Eq.(\ref{TII501}), with $M_s\sim {\cal O}(q^{\ca u})$. Thus the $4(R-S)$ hypermultiplet moduli $q^{\ca u}$ are attracted to $0$, corresponding to the conifold locus in the Higgs branch.

\subsubsection*{Attraction to non Abelian loci}

Near the non Abelian locus the scalar fields relevant to the extremal transition are those in the $SU(N)$-vector multiplet, $X^a$ ($a=1,\dots,N^2-1$), as well as those in the $g$ hypermuliplets in the adjoint of $SU(N)$, $q^{a\ca u}$ ($\ca=1,\dots,g;\ u=1,2,3,4$), and we suppose $g>1$.\footnote{When $g=0$, the pure $SU(N)$ gauge theory has UV freedom, and is Abelian in the IR with the gauge group $U(1)^{N-1}$. This situation can be regarded as an example of the conifold case with $S=N-1$ and $R=0$. For $g=1$, the $SU(N)$-vector multiplet and the only hypermultiplet in the adjoint representation combine into an $\cn=4$ gauge sector. This case is conformal and is already dealt with in Sec.\ref{hetTI}.} The non Abelian loci can be parameterized as $X^a=0=q^{a\ca u}$. Expanding in powers of $X^a$ and $q^{a\ca u}$, imposing $SU(N)$ isometry, we obtain the bosonic part of the supergravity action to the lowest order \cite{Liu:2011nw}:
\begin{align}
    S\!&=\!{1\over \kappa^2_{(4)}}\!\int  d^4x\,\sqrt{-g}\,\Big[\demi R-l^2\nabla X^a\nabla \b X^a -\nabla\!\mfs Q^{a\dagger}_{\ca} \nabla\! \mfs Q^a_{\ca} \nnR &\ \ -g_c^2\, e^{\ck_{V}}\big\{ l^2[X,\b X]^2 + 4 [\b X, q^{\ca u}]^a[q^{\ca u},X]^a+l^{-2} \vec D^a\cdot\vec D^a\big\}\Big], \label{TII6}
\end{align}
where $l$ is a nonzero constant. The $SU(2)_{\car}$ doublet $\mfs Q^a_{\ca}$ and the D-term $\vec D^a$ are defined as
\begin{align}
    \mfs Q^a_{\ca}=\left(\!\!\begin{array}{c} -q^{a\ca 2}+i\,q^{a\ca 1}\\ q^{a\ca 3}+i\,q^{a\ca 4}\end{array}\!\!\right),\ \ \ \vec D^a=\sum_{\ca, b,c}i\, f^{abc} \mfs Q^{b \ca \dagger} \vec \sigma \mfs Q^{c \ca},\label{TII7}
\end{align}
where $f^{abc}$ are the structure constants of $SU(N)$. The action (\ref{TII6}) thus describes an $\cn_4=2$ $SU(N)$ super Yang-Mills field theory formally coupled to gravity. We show that moduli can be attracted to the non Abelian locus from either the Coulomb branch or the Higgs branch.

\noindent $\bullet$ In the Coulomb branch, corresponding to the compactification on $M$, all Cartan components $X^{\h a}$ and $q^{\h a\ca u}$ ($\h a=1,\dots,N-1;\ \ca=1,\dots,g;\ u=1,2,3,4$) acquire nonzero VEV's, while all the non Cartan components vanish. Therefore $X^{\h a}$ form $N-1$ of the $h_{11}(M)$ \Ka moduli, while $q^{\h a\ca u}$ form $4g(N-1)$ of the $4h_{12}(M)+4$ complex structure moduli. The free energy density takes the form of Eq.(\ref{TII501}) with $M_s\sim {\cal O}\big(X^{\h a}, q^{\h a \ca u}\big)$. Therefore the non Abelian locus where $X^{\h a}$ and $q^{\h a\ca u}$ vanish is the local minimum of the free energy density. By consequence,  $X^{\h a}$ and $q^{\h a\ca u}$ are attracted to the non Abelian locus.

\noindent $\bullet$ In the Higgs branch, corresponding to the compactification on $M''$, $q^{a \ca u}$ ($a=1,\dots,N^2-1;\ \ca=1,\dots,g;\ u=1,2,3,4$) have nonzero VEV's satisfying the constraint $\vec D^a=0$ modulo gauge orbits, and they form $4(g-1)(N^2-1)$ of the $4h_{12}(M'')+4$ complex structure moduli. The scalars in the $SU(N)$-vector multiplet $X^a$ vanish. The $SU(N)$-vector multiplet absorbs one hypermultiplet in the adjoint of $SU(N)$ and becomes a long massive vector multiplet. The free energy density is of the form Eq.(\ref{TII501}), with $M_s\sim {\cal O}\big(q^{a \ca u}\big)$. Thus the $(g-1)(N^2-1)$ complex structure moduli $q^{a \ca u}$ are attracted to $0$, corresponding to the non Abelian locus in the Higgs branch.

\subsubsection*{An example: stabilization at intersections of extremal transition loci}

We analyze a 2-parameter example with heterotic dual, where the internal CY space is $M\in\mathbf P^4_{(1,1,2,2,6)} [12](2,128)$. Its mirror is defined by \cite{Klemm:1996kv}
\begin{align}
    x_1^{12}+x_2^{12}+x_3^6+x_4^6 +x_5^2- 12\, \psi\, x_1 x_2 x_3 x_4 x_5-2\, \phi \, x_1^6 x_2^6=0.\label{EXAMP1}
\end{align}
The complex coefficients $\phi$ and $\psi$ are the two \Ka moduli (from the type IIA point of view). This model has at once a conifold locus with $R=S=1$, and an $SU(2)$-non Abelian locus with $g=2$. The latter leads to a Higgs branch corresponding to the CY space $M''\in \mathbf P^5_{(1,1,1,1,1,3)} [2,6](1,129)$. These singular loci are defined by the vanishing of \cite{Klemm:1996kv}
\begin{align}
    \Delta=\Delta_c\times \Delta_s=\big((1-z_1)^2-4z_1^2 z_s\big)\times\big(1-4z_s\big),\label{EXAMP2}
\end{align}
where $\Delta_c=0$ defines the conifold locus, and $\Delta_s=0$ the non Abelian locus, with $z_1=-{1\over 864}{\phi\over \psi^6}$, $z_s={1\over 4\phi^2}$ a reparametrization of the \Ka moduli. The two singular loci intersect at two points: $\big(z_1, z_s\big)=\big({1\over2},{1\over4}\big)$ and $\big({\tiny \infty},{1\over4}\big)$, which are the favored points of moduli stabilization, since there is a maximal number of massless modes at these points. Thus sitting in the Coulomb branch, we can lift the whole \Ka moduli space and $2$ of the $128+1$ quaternionic flat directions in the complex structure moduli space. Also in the Coulomb branch, the heterotic dual compactified on $K3\times T^2$ exists\cite{DualConst}. Therefore we can infer from the type II side the stabilization of the dual heterotic moduli. Especially since the whole vector multiplet moduli space is lifted, the heterotic dilaton, living in a vector multiplet, can be stabilized.

\section{Summary and perspectives}


We have studied moduli stabilization by thermal effects in the cosmological context. The breaking of supersymmetry generates a thermal free energy at one-loop level. The moduli are attracted to its local minima, where extra massless modes appear in the low energy spectrum. These extra massless states can either be perturbative or non perturbative. The scalar masses induced by such thermal effect are time-dependent, which ensures that the universe is radiation dominated at the exit of the intermediate era, so that the cosmological moduli problem does not arise.

Detailed analysis is carried out first to maximally supersymmetric heterotic strings in the weak coupling regime. It is reported for spacetime dimension $D\geq 4$ that all moduli except the dilaton are stabilized at enhanced gauge symmetry points, where the extra massless states are perturbative. Additionally for $D\geq 5$, the dilaton is frozen somewhere in the flat direction, while for $D=4$, it has a logarithmic behavior. Passing to the dual type I picture using the S-duality, one finds that for  $D=4,5$ ($D\geq 7$), all the internal type I moduli can be stabilized in the weak (strong) coupling regime, with the dilaton frozen somewhere in the flat direction. However for $D=6$, where the S-duality map exchanges the heterotic (type I) dilaton with the type I (heterotic) internal volume, the internal volume is frozen in the flat direction and all other moduli including the dilaton are stabilized. The extra massless states are either non perturbative D-string states or perturbative open string states.

Another model studied is the type II strings compactified on CY three-folds. The moduli space admits particular loci where 2-cycles or 3-cycles in the internal CY manifold shrink to zero size, leading to conifold transition or non Abelian gauge symmetry. Extra massless $\cn_4=2$ supermultiplets arise at these loci, inducing local minima to the one-loop free energy. The analysis is based on writing out the full effective action without integrating out the extra light states, so that the action is free of IR divergences. As a result, all type II moduli characterizing the deformation of the shrinking cycles are stabilized. More generally, the favored points in the moduli space are the intersection points of several such loci. An explicit example is given where moduli are stabilized at the intersection of a conifold transition locus and a non Abelian locus, where the entire \Ka moduli space is lifted. This implies in the dual heterotic picture that all vector multiplet moduli are stabilized, including the heterotic dilaton.

More work can be carried out on models with $\cn_4=1$ supersymmetry, for instance, the type II models compactified on generalized CY spaces \cite{geneCY} including fluxes, branes and/or orientifold projections. As mentioned in the introduction, realistic models require a zero temperature spontaneous supersymmetry breaking mechanism, so that the $\cn_4=1$ supersymmetry remains broken at low temperature. Thus it would be of interest to extend the orbifold model results in Refs.\cite{cosmoD} to the context of generalized CY compactifications. Moreover, toroidal compactifications of type II strings in the presence of ``gravito-magnetic'' fluxes lead to thermal models free of Hagedorn-like divergences, and the induced cosmology has no initial singularity \cite{Florakis:2010is}. Therefore we can investigate the implementation of gravito-magnetic fluxes in the (generalized) CY compactifications, hopefully to obtain a theoretical framework giving a full account for both primordial and late-time universe.


\section*{Acknowledgement}

I am grateful to the organizers of the QTS7 conference for the opportunity to present these results. The works reviewed here are in collaboration with John Estes and Herv\'{e} Partouche. I would like to thank also Guillaume Bossard, Pierre Fayet, Albrecht Klemm, and Eran Palti for helpful discussions. This work is supported partially by the contracts PITN GA-2009-237920, ERC-AG-226371, ANR 05-BLAN-NT09-573739, CEFIPRA/IFCPAR 4104-2 and PICS France/Greece, France/USA.

\end{document}